\documentclass{ieeeaccess}
\usepackage{cite}
\usepackage{amsmath,amssymb,amsfonts}
\usepackage{algorithmic}
\usepackage{graphicx}
\usepackage{textcomp}
\usepackage{url}
\usepackage{longtable}
\usepackage{breakurl}
\usepackage{graphicx}
\usepackage{caption}

\def\BibTeX{{\rm B\kern-.05em{\sc i\kern-.025em b}\kern-.08em
    T\kern-.1667em\lower.7ex\hbox{E}\kern-.125emX}}
\begin{document}
\history{Date of publication xxxx 00, 0000, date of current version xxxx 00, 0000.}
\doi{10.1109/ACCESS.2017.DOI}

\title{Disarming Attacks Inside Neural Network Models}
\author{\uppercase{Ran Dubin}\authorrefmark{1,2}, \IEEEmembership{Member, IEEE}
}
\address[1]{Department of Computer Science,\\ Ariel University Ariel, Israel \\ (e-mail: rand@ariel.ac.il)}
\address[2]{Ariel Cyber Innovation Center,\\ Ariel University Ariel, Israel}
\markboth
{Author \headeretal: Preparation of Papers for IEEE TRANSACTIONS and JOURNALS}
{Author \headeretal: Preparation of Papers for IEEE TRANSACTIONS and JOURNALS}

\corresp{Corresponding author: Ran Dubin (e-mail: rand@ariel.ac.il).}

\begin{abstract}
Similar to the revolution of open source code sharing, Artificial Intelligence (AI) model sharing is gaining increased popularity. However, the fast adaptation in the industry, lack of awareness, and ability to exploit the models make them significant attack vectors. By embedding malware in neurons, the malware can be delivered covertly, with minor or no impact on the neural network's performance. The covert attack will use the Least Significant Bits (LSB) weight attack since LSB has a minimal effect on the model accuracy, and as a result, the user will not notice it. Since there are endless ways to hide the attacks, we focus on a zero-trust prevention strategy based on AI model attack disarm and reconstruction. We proposed three types of model steganography weight disarm defense mechanisms. The first two are based on random bit substitution noise, and the other on model weight quantization. We demonstrate a 100\% prevention rate while the methods introduce a minimal decrease in model accuracy based on Qint8 and K-LRBP methods, which is an essential factor for improving AI security.
\end{abstract}


\titlepgskip=-15pt

\maketitle

File-based malware remains a favored tool for hackers, allowing them to swiftly introduce and hide malicious code within seemingly benign files. Microsoft Office documents and Adobe PDFs are mainly targeted due to their widespread daily use. Upon opening a malicious file, the concealed malware instantly activates. Many of these file-based malware prove to be challenging to detect. As new zero-day vulnerabilities arise, and one-day vulnerabilities, though known, retain their effectiveness, the threat becomes even more pronounced.
Furthermore, traditional attacks, like macros, continue to pose a threat \cite{blackberryMacro}. Therefore, conventional detection methodologies might often need help to spot them \cite{file-malware}. Even though the issue of file-based malware is widely recognized, \cite{avcompare} reveals an alarmingly low online detection rate of 96.3\%. Low detection rates underline the urgent need for improved solutions. Recent research indicates that even advanced detection tools, such as next-generation antivirus and Endpoint Detection and Response (EDR) systems, occasionally fall short, missing recognized attacks and known detection bypass methods \cite{karantzas2021empirical,edrbypass}.

An important trend in the world of cybersecurity is the rise of open-source malware attacks. Surprisingly, researchers have discovered that cyber-attacks aimed at open-source repositories have increased by a staggering 633\%. In fact, threat actors have recently uploaded a shocking 144,294 phishing-related packages onto open-source package repositories such as NPM, PyPi, and NuGet \cite{opensource-attack}. This is a serious issue that everyone should be aware of in order to protect themselves from potential cyber threats.

Artificial Intelligence (AI) 's increasing popularity and fast adoption in the industry have led to a new type of file-based malware attack using AI model malware. In those types of attacks, an attacker is trying to hide part of the attack inside the model using steganography \cite{wang2021evilmodel,wang2022evilmodel} or use the model file loading as the first step of the attack \cite{pickle-attack}. This highlights the fact that the machine learning model serialization step used to save the model is vulnerable and can be exploited for cyber-attacks. With the rise in prominence of model zoos such as Hugging Face \cite{Huggingface}, PyTorch Hub \cite{pytorchhub}, and TensorFlow Hub \cite{tensorflowhub}, which offer a variety of free state-of-the-art pre-trained models for anyone to download and utilize, malicious attackers find promising ways to attack users. Actors can release free AI models or hijack / alternate existing models before deployment as part of a supply chain attack. Hugging Face tries to prevent malicious model spreading by scanning models with an open source malware scanner \cite{ClamAV} and detecting malicious PyTorch weight serialization (pickle) vulnerability that can lead to code execution when the model is loading\cite{Fickling}.

Recent research proposes concealing malware in Neural Network (NN) models by substituting model weights bits with malware bits. StegoNet \cite{liu2020stegonet} and EvilModel \cite{wang2021evilmodel} propose a Least Significant Bytes (LSB) steganography embedding malware technique. The LSB steganography attack takes into account that common frameworks, such as PyTorch and TensorFlow, use 32-bit floating-point numbers. By modifying the model weights and using LSB steganography that hides the malware code in different LSB sections of the model weights, attackers can hide a malicious payload in the model with minimal impact on the model's performance and avoid antivirus detection \cite{wang2021evilmodel}.

Using steganography to hide malicious code/commands is not new and is used in other domains. For example, GifShell \cite{Gifshell} uses Gif image to hide command and control, and it is reported that there is a growth in steganography attacks in other file types, including video \cite{steg-video}. However, using NN has a significant advantage compared to other file types:
1) Because of the redundant neurons and excellent generalization ability, the modified neural network models can still maintain the performance in different tasks without causing abnormalities \cite{wang2021evilmodel}. 2) The size of modern models can be used to hide large-size malware. 3) An attacker can embed the malware when the model is saved or infect the model updates. 4) Other formats are well known, and organizations and governments use CDR to remove the threats. For example, NSA published guidelines for sanitizing attack vectors in Microsoft RTF  \cite{NsaRtfSanitization}. However, this is the first work that presents how to disarm LSB steganography attacks hidden in NN model weights. 

This work focuses on novel zero-trust prevention that neutralizes hidden malware attacks in the model rather than detecting malware or employing steganography attacks. This distinguishes it from previous works.

Since detection is not enough from the lesson learned about the state-of-the-art malware detection rate, we propose to use a zero-trust prevention paradigm called Content Disarm and Reconstruction (CDR) \cite{dubin2023content}, but that will focus on deep learning models. In this work, for the first time, we suggest disarming and reconstructing the Neural Network (NN) weight. We propose two novel solutions for disarming steganography attacks embedded in LSB model weights. The first random bit substitution noise prevents attackers from successfully extracting the hidden malware from the model. The second uses a model optimization technique called model quantization \cite{nvidiaquant} to alert the model weight and prevent the attackers from extracting the malware code. The generic prevention solution can be used on any deep learning architecture and defend against any data-hiding strategy in model weights.

The contributions of this paper are summarized as follows:
\begin{itemize}
\item We propose an open source framework \cite{ai-model-cdr-git} to embed malware in the NN model weights and hide malware in the Least Significant Bytes (LSB).
\item We propose three LSB model attack algorithms.
\item  For the first time, we suggest disarming and reconstructing the Neural Network (NN) weight, and we measure the success of different disarm methods against different stenography malware attack strategies.
\item We suggest two algorithms for AI model disarm based on the random bit substitution noise. The methods are designed to disrupt the attacker's ability to extract the malware hidden by steganography attacks in LSB model weights. The generic prevention solution can be used on any NN architecture and defend against any data hiding strategy in model weights. 
\item We evaluate the model weight optimization method called model quantization as a possible CDR algorithm. Quantization is a method known to reduce model complexity and size, and it is based on converting the neuron weight from float 32 bits to 8-bit int. Therefore, it is also an alternative method for a CDR algorithm.
\item We evaluate the impact of embedded malware on the model and the effect of CDR on malicious and non-malicious models and measure 
the performance of CDR algorithms over seven well-known AI  models.
\item We release our steganography model attacks and all CDR algorithms as open-source code \cite{ai-model-cdr-git}.
\item We discuss the advantages and limitations and address future research on the subject.
\end{itemize}

The remainder of this paper is structured as follows. Section \ref{Related Works} describes the related works. Section \ref{Methodology} presents the Methodology for embedding the malware (steganography) and presents three steganography attacks.
Section \ref{CDR Algorithm} presents two methods based on Random Bit Substitutions and one based on model weights quantization.
Section \ref{Evaluation} summarizes our evaluation and results. Section \ref{Discussion} summarizes the future of AI model attacks, its limitation, and future work. Finally, conclusions are summarized in Section \ref{Conclusions}.

\section{Related Works}
\label{Related Works}

\begin{figure}[h]
\centering
\includegraphics[scale=0.6]{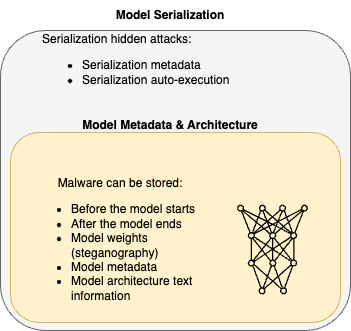}
\caption{Simplified illustration of model file structure composed of serialization format; inside, we have the model architecture, metadata, and weights}
\label{fig_DeepCDR}
\end{figure}

CDR technologies are typically deployed at the organization network/ file upload server entrance or as a service for scanning all incoming files to the organization using agents installed on the device \cite{dubin2023content, dubin2023content-pdf}. CDR will receive the file, separate the file into its discrete components and handle each discrete component based on the component's possible attack vectors. For example, Fig. \ref{fig_DeepCDR} illustrates the model file structure in a simplified manner since different serialization file formats are built differently, but the general concept remains the same. The model serialization may contain metadata that can hide malicious malware or be exploited to run evasive code \cite{pickle-attack} automatically. Inside the serialization, the model is saved. The model contains the model metadata, model architecture that contains textual representation, and model weights. Each part of the model may be used to hide malicious code. CDR relies on the file type format understanding to disarm and reconstruct the file so it will be usable and secure.

As far as we know, this work is the first to suggest CDR for NN models. Works that proposed detection of AI models exploitation/attacks \cite{model-fickling} or detection of malicious stenography \cite{alodat2022detection} are beyond this work scope. This work focuses on zero-trust prevention using CDR methodology regardless of the ability to detect a threat. The proposed CDR solution is done on every received NN model.

Sim et al.~\cite{sim2018defending}, and Sunshine et al. ~\cite{sunshine2021rise} present the need for CDR for different file types but do not discuss how to build and validate the technology and do not present its effectiveness. Han et al.~\cite{han2019secure} present CDR technology for different file types by saving the original document and converting it to a JPEG image file. The advantage of the method is simplicity and security, but the method output is always a JPEG image and not a document. This work focuses on receiving a model as an input, and disarming and reconstructing it as the same model format type without the attacker's ability to extract the hidden malware/data. As a result, the model is fully functional and has similar characteristics to the original model.
 Other CDR-related works focus on PDF privacy content and sensitive data \cite{hiddenPdfdata, Aura-pdf-identify-info, Feng-pdf-privacy, Garfinkel, Sanchez-sanitization}.
 Our previous works suggested CDR for Microsoft RTF file-format \cite{dubin2023content} and PDF file-format \cite{dubin2023content-pdf} against malware attacks. Recent work \cite{belkind2023open} proposed the most similar related work that focused on CDR for images against malware and steganography attacks. However, the current research methodology, evaluation methods, and attacks differ from previous works.

While the NSA has released guidelines for Microsoft RTF content sensitization/disarm \cite{NsaRtfSanitization} and PDF sensitization \cite{NSA-hiddendata2015}, there are, however, no definitions for disarming AI models. This work is the first step toward this goal. As a result, we share our steganography malware attacks and CDR code \cite{ai-model-cdr-git} with the community. The goal of this work is to raise the awareness of the research community against malware steganography hidden inside models and provide zero-trust solutions.

\section{Methodology}
\label{Methodology}
This section introduces the attack, disarm, and reconstruction flow methodology in Section \ref{Workflow}. Then we discuss the steganography attack design in Section \ref{Neural Network Steganography}.

\subsection{Workflow}
\label{Workflow}
\begin{figure*}[h]
\centering
\includegraphics[scale=0.5]{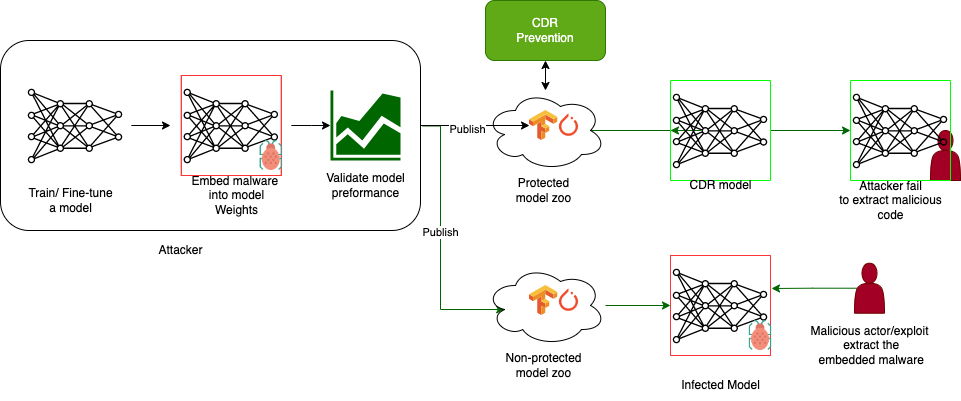}
\caption{Methodology overview from creating the attack and what happens when CDR prevention is used/not used.}
\label{fig_attack_disarm}
\end{figure*}

Fig. \ref{fig_attack_disarm} illustrates the overall workflow from the beginning of the attack. First,  the attacker creates a new model by downloading an existing model or creating a new model. Then, the attacker embeds the malicious code inside the weights in the next step. The attacker's goal is that the embedded malware will not be detected by antivirus or other detection mechanisms. Once the user loads the model, he will not notice a significant change in the model's performance. Note that the attacker can decide to freeze the weights he attacked and re-train the rest of the weights to improve the model's accuracy.

 The attacker can decide with what strategy to embed the malware. The attacker can choose which neurons to hide the malware and how many bits, and the locations of the bits he would want to use. The attacker knows the embedding algorithm, but the defender does not know if the model contains malicious code or where it is embedded. It is harder to detect the embedding in larger models. After the embedding, the attacker can evaluate the model's performance and decide if to release the current attacked model or change his attack tactics to improve the model's accuracy before releasing the attack.

In the next step, the attacker can upload it to a model zoo such as Tensorflow/PyTorch Hub. If the Hub employs the suggested CDR solution, the malware will be disarmed, as presented in Section \ref{Evaluation}. As a result, the malicious code waiting to extract the next phase of the attack hidden in the model will fail. On the other hand, when the Hub is not using CDR, its detection may fail to spot the malware \cite{hiddenlayer}, and the attack will be successful. Therefore, this work does not focus on exploiting /extracting the attack code from the NN, only disarming it and preventing the attacker from extracting the steganography malicious code. As far as we know, there is no detection or prevention mechanism to protect sharing of AI models. The current model Zoo security solutions checks for known serialization exploitation (Pickle) or scan for malware\cite{Huggingface}. However, currently, steganography attacks are undetected.

\subsection{Neural Network Steganography}
\label{Neural Network Steganography}

Typical NN networks consist of: an input layer, hidden layers, and an output layer. The input layer receives external signals and passes the signals to the hidden layer of the neural network through the input layer neurons. The hidden layer neuron receives the incoming signal from the neuron of the previous layer with a certain connection weight. Then, it outputs it to the next layer after adding a certain bias.
 Neurons in the hidden layer can be formulated with weight $w_{i}$ for each input signal $x_{i}$ from the previous layer. Assume that all inputs of the neuron $x = (x_{1}, x{2}, ..., x{n})$, and all connection weights  $w = (w_{1}, w{2}, ..., w{n})$, where $n$ is the number of input signals (the number of neurons in the previous layer). The neuron receives input signal $x$ and calculates $x$ with the weights $w$ by matrix operations with the added bias $b$. As a result, the output of the neuron is calculated by $y = f(wx,b) = f(\sum_{i=1}^{n} w_{i}x_{i},b)$.
The last layer (output layer) receives the incoming signal from the previously hidden layer and processes them to estimate the NN output.

We can conclude that each neuron contains $n$ parameters plus bias which means $n+1$ total parameters. Therefore, NN consists of $m$ neurons with $m(n+1)$ parameters, and in this work, we focus on TensorFlow and PyTorch frameworks where each neuron contains a 32-float parameter (4 bytes). To summarize, the size of each layer is $4m(n+1)$.

\begin{figure}[h]
\centering
\includegraphics[scale=0.4]{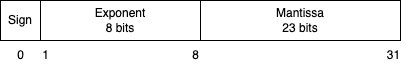}
\caption{Float 32-bit number format.}
\label{fig_float_format}
\end{figure}

An attacker who wants to exfiltrate sensitive data or embed malicious malware in NN can use steganography to hide the data inside the NN weights. The attacker can use steganography based on different hiding strategies and repeat the inverse process to extract the hidden data. The float neuron value can be positive or negative and conforms to IEEE standard \cite{float_ieee}.
Fig. \ref{fig_float_format} presents a 32-bit floating number format.

\begin{figure}[h]
\centering
\includegraphics[scale=0.4]{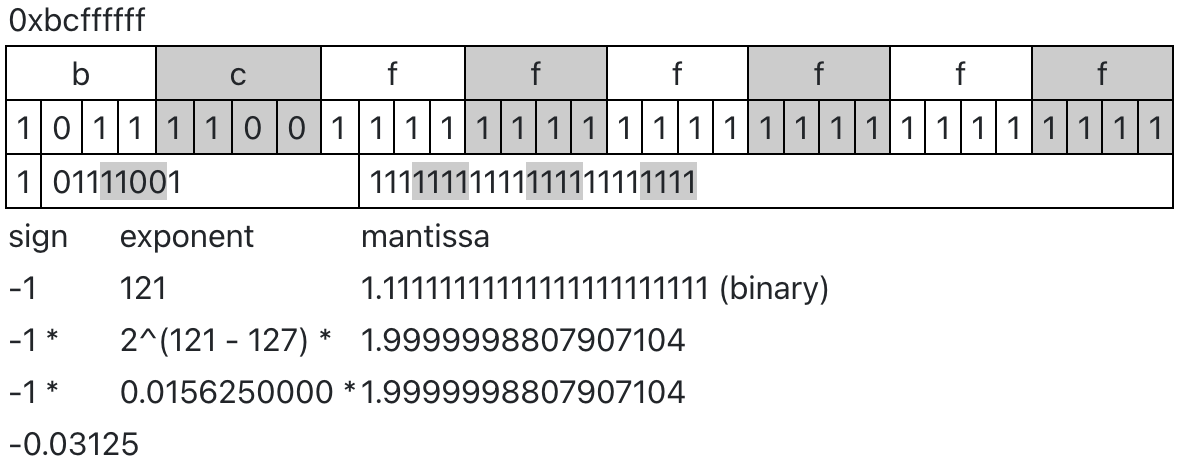}
\caption{Conversion of $0x3C000000$ to float number and explanation on how it is calculated \cite{float_hex_con} }
\label{fig_conv_3c}
\end{figure}

\begin{figure}[h]
\centering
\includegraphics[scale=0.4]{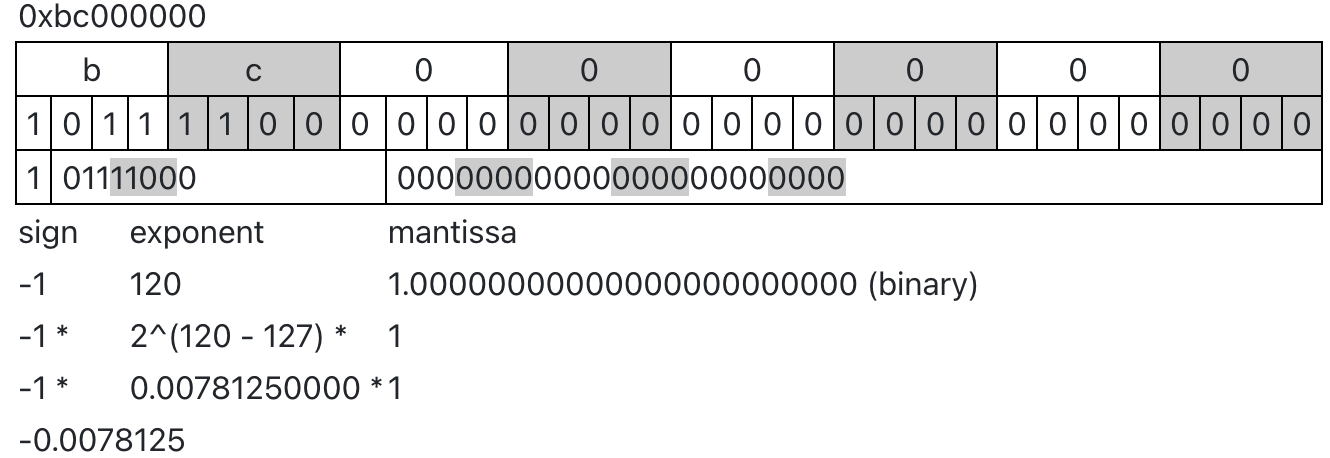}
\caption{Conversion of $0x3CFFFFFF$ to float number and explanation on how it is calculated \cite{float_hex_con}.  }
\label{fig_conv_3cff}
\end{figure}

The float is represented first by the sign with 1 bit, Exponent (2nd -9th bits), and Mantissa (10th- 32nd bits). Let's investigate the float (little-endian representation). We can see the effect of the mantissa over the number representation. In this discussion, we will use a number from the model weights. For example, Fig. \ref{fig_conv_3c} shows a float number hex value of $0x3C000000$, and the maximum modification if we substitute all bits $0x3CFFFFFF$ is represented in Fig. \ref{fig_conv_3cff}. The absolute difference between them is 0.0234375, which is relatively small for the maximum substitution. If instead of $0x3CFFFFFF$, we will use $0x3C0000FF$, the absolute difference will be 0.00000024. Considering the large numbers of neurons in modern NN networks, this gives us the ability and diversity to hide malicious code that is hard to detect and may not affect the network.

\subsection{LSB Steganography Substitution Attacks}
\label{Substitution_attacks}
This section will review the three types of attacks we will use as a baseline for evaluation. From the above section, we are focused on substitution neuron weight LSB bits attack \cite{wang2021evilmodel} since it has shown promising results, stealthiness, and minimal decrease in model performance.

In this section, we will use, the ResNet-101 \cite{he2016deep} as a base example to describe the attack capabilities under the different attack numbers of substitution bits we will use and how much data we can hide in the network as a result. For this network, we will attack the 104 Conv2D layers existing in the network, which contains 42,394,816 neurons we can use. Table. \ref{tab:model_summary} will summarize the attack capabilities for all attacked networks investigated in this work.
\begin{itemize}
\item \textbf{Full Mantissa LSB Attack(FMLA)} - the entire Mantissa LSB float bits are replaced in this attack. If we do not have additional data to embed, we finish and do not change the rest of the NN. Fig. \ref{fig_resnet_conv2d_embed} illustrates per Conv2D layer index how much data we can embed with 23-bit substitution per layer. In total, for ResNet-101, we can embed up to 116 MB of malware data.
\begin{figure}[h]
\centering
\includegraphics[scale=0.4]{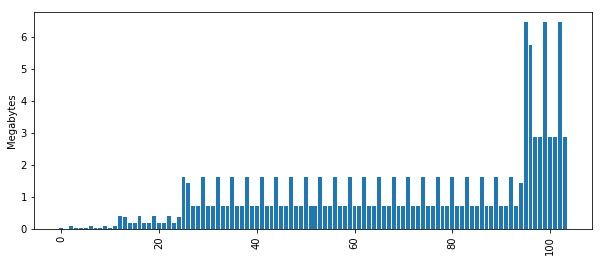}
\caption{An illustration showing the storage capacity of ResNet101 model with 23 bits from each float value. The x-axis indicates the Conv2D layer number, while the y-axis shows the amount of data that can be embedded in megabytes.}
\label{fig_resnet_conv2d_embed}
\end{figure}

\item \textbf{Half Mantissa LSB Attack (HMLA)} - Attacking only the last 12 LSB bits. We can embed up to 60 MB of malware data when using the ResNet-101 model.
\item \textbf{Half Byte LSB Attack (HBLA)} - Using only the last 4 bits of the Mantissa to hide data. In total, we can embed up to 20 MB of data.
\end{itemize}

\begin{table*}[]
\centering
\caption{Summary of the seven models we evaluated. We assessed each model's original size, accuracy, number of neurons, and the number of megabytes it can hide for each type of steganography attack.}
\label{tab:model_summary}
\begin{tabular}{|l|l|l|l|l|l|l|}
\hline
Net       & Size {[}MB{]} & Accuracy {[}\%{]} & \#Neurons  & FMLA & HMLA & HBLA \\ \hline
ResNet101 \cite{journals/corr/HeZRS15} & 170.45        & 75.84            & 42,394,816 & 116          & 60           & 20           \\ \hline
Vgg19 \cite{simonyan2014very}     & 548.14        & 74.218            &    20,018,880        &      54.0        &     28.0         &    9.0          \\ \hline
Vgg16 \cite{simonyan2014very}     & 527.87        & 69.858             &     14,710,464       &    40.0          &    21.0          & 7.0             \\ \hline
Inception \cite{szegedy2015going} & 103.81        & 70.062            &   24,307,040         &      66.0        &    34.0          &    11.0          \\ \hline
ResNet50 \cite{journals/corr/HeZRS15}  & 97.75         & 74.67             &  23,454,912          &     64.0         &   33.0           &  11.0            \\ \hline
ResNet18 \cite{journals/corr/HeZRS15}  & 44.66         & 67.876            &   11,166,912         &        30.0      &      15.0        & 5.0             \\ \hline
Mobilenet \cite{howard2017mobilenets} & 13.55         & 72.028            &    2,942,472        &     8.0         &     4.0         &      1.0        \\ \hline
\end{tabular}
\end{table*}
Table. \ref{tab:Ransomware_attack} illustrates FMLA, HMLA, and HBLA attacks using the Avcrypt Ransomware Portable Executable (PE) file. We can observe that FMLA attacks that modify the entire LSB and, more significantly, the most significant LSB bit (the first Mantissa) cause significant degradation in the model performance, meaning an attacker who uses such an attack is risking possible exposure once the model is validated. The original accuracy is found in Table \ref{tab:model_summary}. However, using HMLA and HBLA in most cases is unnoticeable, as we can see for the Inception model or Mobilenet. The minimal effect is because the attack hides the malware in the most insignificant bits that do not affect float-32. Therefore, attackers will prefer to attack using those algorithms. In some cases, we can observe that FMLA attacks, such as Mobilenet, eradicate the model, while models like VGG16 are more resilient.

\begin{table}[]
\centering
\caption{Models accuracy after an attack using the Avcrypt ransomware file (md5: 248144f924d49b37312da171f14f4131) with a size of 3.1 MB.}
\label{tab:Ransomware_attack}
\begin{tabular}{|l|l|l|l|l|l|l|}
\hline
Net     & FMLA & HMLA & HBLA \\ \hline
ResNet101  &     6.058      &   75.846         &   75.84
     \\ \hline
Vgg19        &      55.828      &    70.14      &     70.134
     \\ \hline
Vgg16      &   61.856      &   69.848         &   69.858
   \\ \hline
Inception       &   32.936     &  70.07         &  70.082
      \\ \hline
ResNet50          &    3.942      &   74.676        &   74.67
       \\ \hline
ResNet18         &      10.278    &    67.874       &    67.876
      \\ \hline
Mobilenet       &     0.084    &    72.04      &    72.028
     \\ \hline
\end{tabular}
\end{table}

Table \ref{tab:vbscript_attack} illustrates a scenario where a small-size visual basic script malware is used instead of PE Ransomware as previously shown. The lower file size results in fewer neurons being affected during the attack, and each attack method results in a lower accuracy decrease because fewer weights were changed. For example, there was no accuracy degradation for Resnet101 with the HBLA attack strategy. In the remainder of this paper, we will concentrate on the results of the ransomware attack. Based on our experiments, the insights across various attacks are consistent. However, we will specifically highlight the ransomware use case due to space constraints.
\begin{table}[]
\centering
\caption{The model's accuracy after being attacked using a Visual Basic script with a size of 176 KB (md5: 1f63c855ebe8adfdedec5d4384582292).}
\label{tab:vbscript_attack}
\begin{tabular}{|l|l|l|l|l|l|l|}
\hline
Net     & FMLA & HMLA & HBLA \\ \hline
ResNet101  &    56.42
       &      75.848
      &      75.84      \\ \hline
Vgg19        &     64.274
       &    70.136
      &      70.134     \\ \hline
Vgg16      &    63.502
     &     69.852
      & 69.858          \\ \hline
Inception       &  62.838
   &   70.068
     &   70.082       \\ \hline
ResNet50          &   54.722
   &     74.674
    & 74.67          \\ \hline
ResNet18         &   46.536
    &   67.882
     & 67.876          \\ \hline
Mobilenet       &  0.132
    &   72.022
    &     72.028      \\ \hline
\end{tabular}
\end{table}

\section{CDR Algorithms}
\label{CDR Algorithm}

We will evaluate two different zero-trust prevention CDR algorithms based on random bit substitutions illustrated in Section \ref{rbs_defense} and compare them to the model weight quantization method in Section \ref{quantization_algo}.

\subsection{Random Bit Substitutions}
\label{rbs_defense}
Our approach focuses on random bit substitutions to modify neural network model weights, intending to thwart potential attackers from extracting embedded code. This strategy is based on two primary assumptions:
a) While a model might already be compromised, there is no definitive way to ascertain this. Modifying the model may compromise its performance.
b) Detection mechanisms cannot be trusted to alert users or services about breaches consistently.
To counteract potential compromises, we introduce two techniques using CDR methods that disarm and subsequently rebuild the model. This aims to ensure its functionality with minimal degradation in quality. The key is to apply our algorithm to every model (operating on a zero-trust principle) irrespective of any perceived threats. This means, however, that the potential slight reduction in model performance should always be considered.

\begin{itemize}
\item {Full LSB Prevention (FLP)} - In this CDR algorithm, we replace all Cov2D neurons with 23 random bits to prevent attacks. The FLP method is the most aggressive but has the highest prevention rate of 100\%. However, it has the most impact on the model.
\item {K-LSB Random Bits Prevention (K-LRBP)} - In this method, per $k$ substitution per neuron, we randomly select k bits to replace. We select k values of 10 and 5, and 1 bit. 
\end{itemize}

\subsection{Model Quantization}
\label{quantization_algo}

Model quantization computes and stores tensors at lower bit widths than floating point (32/64) precision. As a result, a quantized model executes some or all of the operations on tensors with reduced precision rather than full precision (floating point) values, leading to a 4x reduction in model size and memory bandwidth. It is also reported that Hardware support for INT8 computations is typically 2 to 4 times faster compared to floating point 32 \cite{modelquant}. In this work, we will evaluate uint8 and int8 quantization methods. Since this paper focuses on LSB attacks, this is equivalent to removing the LSB, and int8 modifies the sign value. Model quantization is used in various model optimization and hardware adaptations \cite{zhang2022comprehensive}.

Floating-point numbers are distributed non-uniformly in the dynamic range, and about half of the representable floating-point numbers are in the interval [-1,1]. However, using an 8-bit integer representation, we can represent only $2^8$ different values, all positives (in the case of 8-bit int). Furthermore, these 256 values can be distributed uniformly or non-uniformly, for example, for higher precision around zero. All mainstream, deep-learning hardware and software use a uniform representation because it enables computing using high-throughput parallel or vectorized integer math pipelines \cite{nvidiaquant}.

Formula \ref{quantization_formula} describes a symmetric quantization of a floating point tensor $x_f$ to an 8-bit representation $x_q$. $Clip$ is a function that clips outliers that fall outside the [-128, 127] interval. Formula \ref{quantization_formula_scale} defines the scale parameter which uses the full range that you can represent with signed 8-bit integers: [-128, 127] where $amax$ (Formula \ref{quantization_formula_amax}) describes the element with the largest absolute value to represent. It is essential to point out the decision to represent using an 8-bit integer/float, a Clipping function (\ref{quantization_formula}), and the error introduced by the rounding operation, which may decrease the model accuracy.
\begin{equation}
\label{quantization_formula}
x_q = Clip(Round(\dfrac{x_f}{scale}))
\end{equation}

\begin{equation}
\label{quantization_formula_scale}
scale = \dfrac{(2*amax)}{256} 
\end{equation}

\begin{equation}
\label{quantization_formula_amax}
amax = max(abs(x_f))
\end{equation}

To address the effects of losing the precision of the model weights, various quantization techniques have been developed \cite{nvidiaquant,gholami2021survey}. These techniques belong to one of two categories: post-training quantization (PTQ) or quantization-aware training (QAT). Traditional PTQ is performed after a high-precision model has been trained by quantizing the weights and updating the activation function distributions using a subset of the model dataset. However, in the scope of this work, the CDR model receives only the model from the model zoo and doesn't have the training/ testing dataset. Therefore the quality of the model is reduced even more due to the need for more ability to optimize the model. Similarly, this is why PTQ can't be done in our scenario since we are not training the model and only receiving the model. In this work, we will compare 8-bit sign quantization and name it in our result as $Qint8$.

\section{Evaluation}
\label{Evaluation}
This section will evaluate the disarm algorithms' effect over the original models without steganography attacks.
Table \ref{tab:clean_models_CDR} summarizes the effect of CDR on the original model without attacking it. CDR is applied to every received model, and this evaluation aims to understand the impact of CDR algorithms on the original model without prior detection as an assumption (zero-trust). We compare FLP, 1-LRBP, 5-LRBP, 10-LRBP, and Qint8. We can observe that most models crash completely when using FLP. However, using the K-LRBP strategy or Qint8 provided similar results to the original detection, which means they are applicable to protect the models. In ResNet18, after Qint8, the model shows a minimal improvement in accuracy which is negligible. It is important to emphasize that the same test datasets matching each model were used.
\begin{table*}[]
\centering
\caption{The effect of CDR over the original models without attacking the models.}
\label{tab:clean_models_CDR}
\begin{tabular}{|l|l|l|l|l|l|l|}
\hline
Net      & Accuracy {[}\%{]} & FLP  & 1-LRBP & 5-LRBP & 10-LRBP & Qint8 \\ \hline
ResNet101        & 75.84            & 4.354	& 75.84	& 75.84&	75.842 & 75.84
           \\ \hline
Vgg19           & 74.218            &    47.89 &	70.134 &	70.134 &	70.136 & 70.108
         \\ \hline
Vgg16            & 69.858             &     52.68	&69.858	&69.858&	69.858 & 69.818
            \\ \hline
Inception       & 70.062            &   0.252 &	70.082&	70.082&	70.08 & 70.086
       \\ \hline
ResNet50         & 74.67             &  6.66&	74.67&	74.67&	74.668
  & 74.644         \\ \hline
ResNet18          & 67.876            &   13.118	&67.876&	67.876&	67.882 & 67.884
             \\ \hline
Mobilenet        & 72.028            &   0.116	&72.028&	72.028&	72.034 & 72.022
        \\ \hline
\end{tabular}
\end{table*}

Table \ref{tab:HBLA_model_attack_summary} shows the HBLA (4-bit) model attack and CDR results for each method. FLP constantly modifies the most significant bit in the LSB and always produces unusable results. However, the K-LRBP method with K equal to 1, 5, and 10 bits provides excellent results that match the original accuracy (Table \ref{tab:clean_models_CDR}) for ResNet101, VGG16, Inception, ResNet18, and Mobilenet. There is a difference with VGG19, leading to an accuracy reduction of almost 4\%. However, the rest of the models act similarly. Interestingly, in some cases, 1-LRBP is better than Qint8, as we can see in Mobilenet, while in ResNet18, we can see the opposite. The CDR in all the results provided 100\% security and the malware inside the model could not be extracted.
\begin{table}[]
\centering
\caption{The effect of CDR over the original models with attacking the models using the Ransomware file and HBLA attack strategy.}
\label{tab:HBLA_model_attack_summary}
\begin{tabular}{|l|l|l|l|l|l|l|}
\hline
Net & FLP & 1-LRBP & 5-LRBP & 10-LRBP & Qint8 \\ \hline
ResNet101 & 1.366 & 75.84 & 75.84 & 75.842 & 75.842 \\ \hline
VGG19 & 47.902 & 70.134 & 70.134 & 70.134 & 70.112 \\ \hline
VGG16 & 47.462 & 69.858 & 69.858 & 69.858 & 69.838 \\ \hline
Inception & 0.228 & 70.082 & 70.082 & 70.076 & 70.086 \\ \hline
ResNet50 & 5.114 & 74.67 & 74.67 & 74.67 & 74.644 \\ \hline
ResNet18 & 13.786 & 67.876 & 67.876 & 67.876 & 67.884 \\ \hline
Mobilenet & 0.114 & 72.028 & 72.03 & 72.032 & 72.022 \\ \hline
\end{tabular}
\end{table}

Table \ref{tab:HMLA_model_attack_summary} summarizes the result of HMLA (12-bit attack) and how the CDR methods disarm the effect over the model's accuracy. Similarly to the HBLA attack prevention, we see consistent results that slightly degraded the model accuracy since the attack used  12-bit and not 4-bit, as in the previous example. However, the CDR managed to disarm all malicious content, and if we compare the accuracy of the model before the CDR and after the attacks, as can be seen in Table \ref{tab:Ransomware_attack}, the results are very similar but with added security. For example, Resnet101's original accuracy is 75.84\%, and after the HMLA attack, the accuracy is  75.846 \%, while after CDR, the accuracy is 75.84\% with 10-LRBP and similarly 75.85\% with Qint8. Therefore, there is no significant reduction in the model's accuracy with CDR.

\begin{table}[]
\centering
\caption{The effect of CDR over the original models with attacking the models using the ransomware file and HMLA strategy}
\label{tab:HMLA_model_attack_summary}
\begin{tabular}{|l|l|l|l|l|l|l|}
\hline
Net & FLP & 1-LRBP & 5-LRBP & 10-LRBP & Qint8 \\ \hline
ResNet101  & 2.044 & 75.846 & 75.846 & 75.844 & 75.85 \\ \hline
VGG19  & 56.568 & 70.14 & 70.14 & 70.136 & 70.138 \\ \hline
VGG16  & 46.534 & 69.848 & 69.848 & 69.854 & 69.84 \\ \hline
Inception & 0.414 & 70.07 & 70.07 & 70.062 & 70.072 \\ \hline
ResNet50 & 8.832 & 74.676 & 74.676 & 74.674 & 74.636 \\ \hline
ResNet18 & 12.204 & 67.874 & 67.874 & 67.876 & 67.88 \\ \hline
Mobilenet & 0.106 & 72.04 & 72.04 & 72.036 & 72.022 \\ \hline
\end{tabular}
\end{table}

Table \ref{tab:FMLA_model_attack_summary} presents the CDR outcomes following the FMLA attack strategy, which employs all 23 LSB bits. As shown in Table \ref{tab:Ransomware_attack}, the FMLA attack impacts the most significant LSB, consistently compromising the model's functionality. The CDR results, as anticipated, closely mirror the initial post-attack outcomes. Given these outcomes, attackers will likely avoid this approach since it lacks subtlety. In contrast, the HMLA and HBLA attacks do not significantly alter the model's accuracy, making them more covert and appealing options.
\begin{table}[]
\centering
\caption{The effect of CDR over the original models with attacking the models using the ransomware file and FMLA strategy}
\label{tab:FMLA_model_attack_summary}
\begin{tabular}{|l|l|l|l|l|l|l|}
\hline
Net & FLP & 1-LRBP & 5-LRBP & 10-LRBP & 8-bit int \\ \hline
ResNet101 & 4.15 & 6.058 & 6.058 & 6.06 & 6.052 \\ \hline
VGG19 & 50.826 & 55.828 & 55.828 & 55.828 & 55.832 \\ \hline
VGG16 & 49.132 & 61.856 & 61.856 & 61.856 & 61.846 \\ \hline
Inception & 0.218 & 32.936 & 32.936 & 32.94 & 32.944 \\ \hline
ResNet50 & 6.176 & 3.942 & 3.942 & 3.944 & 3.938 \\ \hline
ResNet18 & 11.142 & 10.278 & 10.278 & 10.278 & 10.336 \\ \hline
Mobilenet & 0.078 & 0.084 & 0.084 & 0.086 & 0.084 \\ \hline
\end{tabular}
\end{table}

\section{Discussion}
\label{Discussion}
This study introduces an open-source approach for executing steganography attacks on AI models. Additionally, we present a zero-trust prevention strategy, which effectively mitigates these attacks without the need for malware detection. Leading AI systems and anti-virus tools currently struggle to identify steganography-based malware. While a significant volume of prior research has honed in on image steganalysis \cite{muralidharan2022infinite}, there is an emerging necessity to pivot attention toward AI models. Given the soaring popularity of AI models and Large Language Models (LLM), there is potential for malevolent actors to disseminate deceptive models crafted for malware distribution.
Our research focuses on a steganography attack prevention mechanism via CDR that guarantees complete security. Such a mechanism ensures that attackers cannot retrieve their embedded malware from the model while only slightly affecting its confidence.
For subsequent research, attention should be directed toward neutralizing other components of the AI model. As depicted in Figure \ref{fig_DeepCDR}, this includes aspects such as model metadata, its architecture, and serialization techniques \cite{pickle-attack}. Past studies have evidenced that these components can be manipulated to launch attacks when a model is loaded into a computer's memory. By amalgamating all these protective methods, one can fortify an AI model to be fully resilient against a broad spectrum of threats.

\section{Conclusions}
\label{Conclusions}
In our study, we examined three distinct attack strategies: FMLA (23-bit), HMLA (12-bit), and HBLA (4-bit). While attackers can customize their attack variations, these scenarios were chosen to demonstrate how targeting the most significant LSB can severely degrade model accuracy. In contrast, the impacts of HMLA and HBLA on the model are negligible.
We introduced three CDR methods for mitigation. Two of these are based on Random Bit Substitutions: FLP, which randomizes 23 bits (primarily illustrative to underscore the methodology, given its ineffectiveness), and K-LRBP, which has demonstrated outstanding disarming capabilities. Our third approach, Qint8, leverages model quantization. Both Qint8 and K-LRBP ensured 100\% security in our tests, and a crucial observation was that the model accuracy only witnessed a marginal shift, indicating the inconsequential nature of the change.
VGG19 was notably more susceptible to attacks and CDR, registering an approximate 4\% degradation which was abnormal in our evaluation. The deviations were minimal for other models, primarily when the most significant LSB remained unaltered. Attackers would be wise to sidestep targeting the most significant LSB bit, given its pronounced influence on floating-point calculations. The AI malware security, as described in the Discussion section, is in its first steps; however, the ability of enterprise and personal users to fetch AI models automatically from Model Zoo makes AI model security more critical than ever and has to be improved. The proposed solution provides superior prevention based on CDR with minimal model accuracy decrease for the first time.

\section*{Acknowledgment}
 This work was supported by the Ariel Cyber Innovation Center in conjunction with the Israel National Cyber Directorate in the Prime Minister's Office. This work is under US Provisional Patent Application No. 63/524,681. We thank Ofek Alon for his feedback and for running part of the simulations.

\bibliographystyle{IEEEtran}
\bibliography{refs}

\begin{IEEEbiography}[{\includegraphics[width=1in,height=1.25in,clip,keepaspectratio]{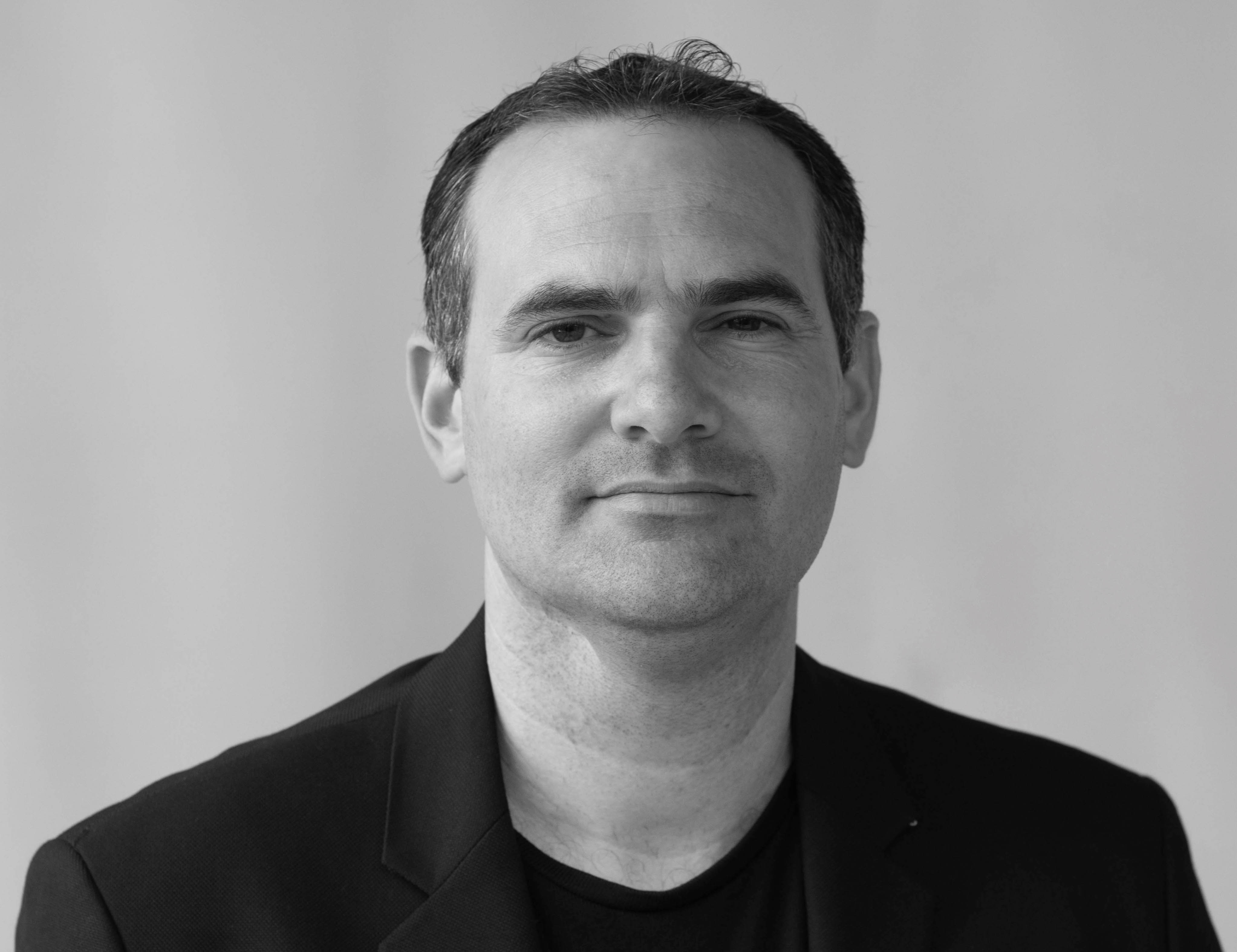}}]{Ran Dubin}
Received his B.Sc., M.Sc., and Ph.D. degrees from Ben-Gurion University, Beer Sheva, Israel, all in communication systems engineering. He is currently a faculty member at the Computer Science Department, Ariel University, Israel. His research interests revolve around zero-trust cyber protection, malware disarms and reconstruction, encrypted network traffic detection, Deep Packet Inspection (DPI), bypassing AI, Natural Language Processing, and AI trust. 
\end{IEEEbiography}

\EOD

\end{document}